\documentclass[10pt]{article}
\usepackage{amsmath}
\usepackage{multicol}
\usepackage{graphicx}
\usepackage{pstricks}
\usepackage{pst-plot}
\usepackage{floatflt}
\usepackage{epsfig}
\usepackage[latin1]{inputenc}

\begin{document}
\vspace{-.5cm}
\title{
IS THE INTERPRETATION OF   DELAYED-CHOICE EXPERIMENTS  MISLEADING?
}
\author{ Pierre Roussel, Iulian Stefan\\
Institut de Physique nucl\'eaire\\
Universit\'e Paris XI, CNRS, IN2P3 \\
F-91406 Orsay Cedex }
\date{}
\maketitle
\begin {abstract}
{\bf The interpretation of  an  experimental realization of Wheeler's delayed-choice gedanken experiment is discussed and called into question. \hspace{4cm} IPNO-DR-07-04}
\end {abstract}

\begin{quotation}
	{\em Once more, we find that nature behaves in agreement with  the prediction of quantum mechanics even  in surprising situations where a tension with relativity seems to appear.}
\end{quotation}

It is this quotation from J.S. Bell \cite{bell87} which brings Vincent Jacques's report \cite{jacq07}   to an end.
What is it about?  In this "Experimental realization of Wheeler's delayed-choice gedanken experiment " 
 a Mach-Zehnder interferometer is used either in a "closed" configuration,  the usual one,  with a phase adjustment leading to the extinction of one photo-multiplicator (PM2), or in an  "open" configuration where the second beam-splitter BS2 has been removed leading to an equal probability  for the triggering of PM1 or PM2, as illustrated in the next  drawing.
\vspace{.8cm}
\begin{pspicture}(10,6.5)
\rput(10,2.5){
\psline{-}(-1,0)(5,0)
\psline{-}(2,2)(6,2)
\psline{-}(2,0)(2,2)
\psline{-}(5,0)(5,3)
\rput(2,0){
\psline[linestyle=dashed]{-}(-0.4,-0.4)(0.4,0.4)
}
\rput(5,0){
\psline{-}(-0.4,-0.4)(0.4,0.4)
}
\rput(2,2){
\psline{-}(-0.4,-0.4)(0.4,0.4)
}
\rput(6,2){
\psline{-}(-0.,-0.3)(0.,0.3)
\psarc(0,0) {.3}{-90}{90}
}
\rput{90}(5,3){
\psline{-}(-0.,-0.3)(0.,0.3)
\psarc(0,0) {.3}{-90}{90}
}
\psplot[linewidth=.01]{-.25}{.75}{1 x -.25 add x -.25 add mul .2 add div .3 mul -.6 add}
\rput [c](.25,1.5){single-photon}
\rput [c](.25,1.1){pulse}
\rput(2.6,.7){BS1}
\rput(4.,3.2){PM1}
\rput(4.,2.7){(prob.=0.5)}
\rput(6.1,1.4){PM2}
\rput(6.1,0.9){(prob.=0.5)}
\rput(2.,-1){\large "Open" configuration.}
}
\rput(1,2.5){
\psline{-}(-1,0)(5,0)
\psline{-}(2,2)(6,2)
\psline{-}(2,0)(2,2)
\psline{-}(5,0)(5,3)
\rput(2,0){
\psline[linestyle=dashed]{-}(-0.4,-0.4)(0.4,0.4)
}
\rput(5,2){
\psline[linestyle=dashed]{-}(-0.4,-0.4)(0.4,0.4)
}
\rput(5,0){
\psline{-}(-0.4,-0.4)(0.4,0.4)
}
\rput(2,2){
\psline{-}(-0.4,-0.4)(0.4,0.4)
}
\rput(6,2){
\psline{-}(-0.,-0.3)(0.,0.3)
\psarc(0,0) {.3}{-90}{90}
}
\rput{90}(5,3){
\psline{-}(-0.,-0.3)(0.,0.3)
\psarc(0,0) {.3}{-90}{90}
}
\psplot[linewidth=.01]{-.25}{.75}{1 x -.25 add x -.25 add mul .2 add div .3 mul -.6 add}
\rput [c](.25,1.5){single-photon}
\rput [c](.25,1.1){pulse}
\rput(2.6,.7){BS1}
\rput(4.4,1.3){BS2}
\rput(4.,3.2){PM1}
\rput(4.,2.7){(prob.=1.)}
\rput(6.1,1.4){PM2}
\rput(6.1,0.9){(no count)}
\rput(2.5,-1){\large "Closed" configuration}
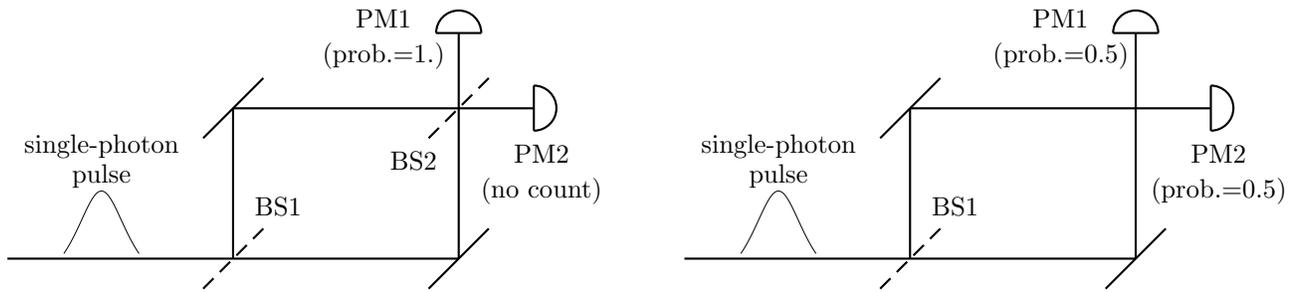
\rput(2.5,-1.5){\large  (with a particular phase adjustment).}
}
\end{pspicture}

It is argued by the author following 
  Wheeler\cite{wheel84} that in the closed configuration each  photon detected at PM1{\em "has arrived by both routes"} whereas in the open configuration {\em "the photon  has travelled only one route".} Because of the short pulse duration in the presented experiment and  since  BS2 can be set or removed while the photon is travelling in the interferometer arm(s), hence after it passed BS1, it results what one could call a "delayed choice effect"  :
\begin{quotation}
{\em "Thus one decides the photon shall have come by one route or by both routes after it has already done its travel"}\cite{wheel84} or else again 
{\em "We, now, by moving the mirror in or out have an unavoidable effect on what we have the right to say about the {\em already} past history of that photon."}
\end{quotation}
In this interpretation, a change is then produced  at the location of BS1 and up to BS2 by  the arbitrary choice of setting in or not BS2.
Here is the delayed-choice effect,  the specific "tension with relativity".  In Wheeler's article consequences could be found back at the time of Big Bang (see footnote  further).
But is this effect or this interpretation true? Does QM  require or predict such a specific effect? We will argue that no, and in order to emphasize that nothing new is invented by the present authors, quotations from the founding fathers of quantum mechanics will be called as often as possible.  

 A tension with relativity is indeed at work  each time  a  "wave packet reduction" occurs within the QM description of a micro-physics experiment and we will examine whether or not it applies here and if something more has to be added. 

 To begin, one has to go back to a long lasting debate between A. Einstein and N. Bohr. The former, as reported by the latter  in 1955\cite{EPS}, had argued that in a double-slit diffraction experiment :

\begin{quotation}
	{\em If in the experiment the electron is recorded at one point A of the plate then it is out of the question of ever observing an effect of this electron at another point (B), although the laws of ordinary wave propagation offer no room for a correlation between two such events.}
\end{quotation}
and indeed one could find the same type of  remark much earlier (in 1927) at the fifth Conseil de Physique Solvay where Einstein commenting this time a single hole diffraction experiment  declared:
\begin{quotation}
{\em If one uses Schrödinger's waves,   the interpretation  of $\vert\Psi ^2\vert$ within a theory of individual events claiming to be complete, implies in my opinion a contradiction with the relativity postulate.}
\end{quotation} 

Does the reported experiment deserve more than Einstein's remarks? It seems that no.

In the "open" configuration (no second beam-splitter BS2) of Wheeler's proposal, nothing more and nothing less is at work than in the Einstein-Bohr gedanken experiment. The space open to the detection of a "diffracted" electron becomes here the two tubes of space following the first beam splitter BS1 in the two arms of the interferometer which are together open to the detection of the photon. The two tubes of space  happen to go across one each other without any particular consequences in this open configuration. Here again " If in the experiment the photon is detected at one point A (wherever is this point) it is out of the question...etc..."
However after detection, the statement that the photon {\bf was} (or travelled) in the arm corresponding to the triggered PM is as illegitimate as would be to draw any trajectory ending at the electron impact in the Einstein-Bohr experiment, or more generally to draw a trajectory between two subsequent position measurements, for, as stated by Schrödinger:

\begin{quotation}
{\em	In general ... a variable \emph{has} no definite value before I measure it ; then measuring it does \emph{not} mean ascertaining the value that it \emph{has} }
\end{quotation} and still less the value that it \emph{had}.

As regards these questions, a central issue is the status of the wave function. It is not the particle which is diffracted or split on the beam splitter mirror, or which travel  "both routes" (or "one route"). The wave is not the object as ascertained by Bohr\cite{bohr35}   :

\begin{quotation}
	{\em The diffraction by the slit of the plane wave giving the symbolic representation of its state will imply etc...}
\end{quotation}

If a strange situation appears since a "symbolic representation" is diffracted by a real slit, nevertheless no doubt for Bohr that the particle is not the wave nor a model of it.
That $\Psi$ is not a model is stressed at length and with more clarity by Schrödinger\cite{schro35}  in its article written (in 1935) in reply to that of EPR (which contains the cat paradox chapter) where chapter 7 is untitled " The $\Psi$-function as Expectation catalog" where he explains -the  well known statistical interpretation of Max Born!- that $\Psi$ is "the means for predicting probability of measurement results". Not only the wave is not  the particle, the object,  but the wave has not to be considered as real when the "reduction" occurs : as written by Bell, a tension with relativity (a contradiction for Einstein),  the same observation as that concerning the  EPR, Einstein Podolsky Rosen, correlations.

Up to the location of the second beam splitter BS2, the probability of detecting a photon is one half all along each arm for each triggered event. If time is considered in the reported particular experimental situation, this probability has a time evolution to match the wave packet propagation.This does not depend on the presence or not of BS2. Of course it is no more true after the place of BS2 for if absent it leaves the sharing of the probability between the two arms unchanged whereas if present it changes  drastically the subsequent wave function (or may do so if the phases are so adjusted) and consequently the detection probabilities.

It must be clearly settled that no change in the wave function is introduced by BS2 between BS1 and BS2.  Assuming on the contrary that the photon \emph{was} in one or the other arm in the open configuration would simply mean that after BS1 the photon is now in a mixture of state, instead of a pure state ("a complete expectation-catalog"," a maximum knowledge state", a  $\Psi$-function in Schrödinger wording) as precisely required and demonstrated by the observed results in the closed configuration. Only a measurement, a wave packet reduction, is able to produce a change from  a pure state to a mixture of states.

The use of Bohr's complementary principle is called in Wheeler's demonstration as well as in Jacques's article:

\begin{quotation} {\em 
Such an experiment supports Bohr's statement that the behavior of a quantum system is determined by the type of measurements performed on it. Moreover, it is clear that for the two complementary measurements considered here, the corresponding experimental settings are mutually exclusive ; that is, BS2 cannot be simultaneously present and absent. }
\end{quotation} 

 Is this call to complementarity  appropriate? This principle (actually more a warning than a principle) stresses that the use of experimental settings corresponding to conjugate variables are exclusive. After a common preparation leading to an initial state and its corresponding (pure state) wave function, the changes brought by the  further instrument will be different depending on the variable to be measured and the actual subsequent measurement results will also be different. This principle does not claim that the initial state is changed as finally suggested by the "delayed choice" experiment or rather its interpretation. No doubt that {\em after} the location of BS2 results will depend in an exclusive way on the presence or not of BS2 but nothing has to be changed before.

If one is now convinced that this "delayed choice effect" is not really required by  QM, why did  J.A. Wheeler invent it? And finally, did really J.A. Wheeler believe himself  in this effect he contributed so much to propagate?  This may be questioned when in the same article one finds:

\begin{quotation}{\em
"No elementary phenomenon is a phenomenon until it is a registered (observed) phenomenon". It is wrong to speak of the "route " of the photon in the experiment of the beam splitter. It is wrong to attribute a tangibility to the photon in all its travel from the point of entry to its last instant  of flight.	}
\end{quotation}

This quotation\footnote {another  quotation concerning a "{\em cosmological delayed choice experiment}" carries the same meaning :
\begin{quotation} {\em 
 "This result, again in a misleading phraseology, 
%{\em (it is this quotation which justify the quotes in our title!)}
 says that "the photon in question came by {\em both} routes. "However, at the time the choice was made whether to put in BS2 or leave it out, the photon in question had {\em already } been on its way for billions of years. It is not right to attribute to it a route. No elementary phenomenon is a phenomenon until it is a registered  phenomenon."}(note : almost a repetition but "observed" is missing!)
\end{quotation} }
 begins with the  so often reported statement which during decades  suggested  a role for the observer's conscience ; but the rest of the quotation    seems much closer to  standard  QM.
Why has this part of Wheeler's article ignored for so long?
If the beautiful experiment reported by Vincent Jacques  and performing Wheeler's gedanken experiment  does not produce  any  "delayed choice effect",  it does exemplify the observation of a wave packet reduction in a very special situation  :  because of the short duration of the light pulse the wave packet is split into two parts  with no connexity,  which  "travel" separately along the spectrometer arms.

\end{document}